\documentclass[aps,prc,preprint,superscriptaddress,10pt]{revtex4-2}
\usepackage{graphicx}  
\usepackage{dcolumn}   
\usepackage{bm}        
\usepackage{amsmath}   
\usepackage{amssymb}   
\usepackage{url}       

\usepackage[colorlinks,citecolor=blue,urlcolor=blue,linkcolor=black,
                   bookmarks=false,hypertexnames=true]{hyperref}

\begin{document}

%
\title{Mapping Dirac--Hartree--Fock approach onto relativistic mean field model}

\author{Stefan Gmuca}

\author{Kristian Petr\'ik}
\affiliation{Institute of Physics, Slovak Academy of Sciences,
SK-845 11 Bratislava 45, Slovakia}

\author{Jozef Leja}
\affiliation{Institute of Physics, Slovak Academy of Sciences,
SK-845 11 Bratislava 45, Slovakia}
\affiliation{Faculty of Mechanical Engineering, Slovak University
of Technology, SK-812 31 Bratislava, Slovakia}

\date{\today}

%
\begin{abstract}
The exchange part of energy density of the linear
Dirac--Hartree--Fock (DHF) model in symmetric nuclear matter is
evaluated in a parameter--free closed form and expressed as density
functional. After the rearranging terms the relativistic mean-field
approach with density-dependent couplings may be recovered with
density dependence coming from the Fock exchange. The formalism
developed, is then extended to the nonlinear DHF approximation with
field self-couplings allowed. The nonlinear self--interactions
result in essential density dependence (medium modification) of effective
couplings that are decoupled from the Fock ones.
\end{abstract}



%
\maketitle


\section{Introduction}
\label{sec:Intro}

Kohn and Sham's density functional theory (DFT) \cite{Hohenberg:1964,*Kohn:1965}
has developed into one of the most successful
approaches ever for a description of the electronic structure of matter.
It provides the great variety of ground--state properties of a system
with the electron density playing the key role.

Inspired by the achievements of the coulombic DFT, there is
an ambition of current nuclear physics to develop the
(energy) density functional for a proper description of exotic
nuclei and equation--of--state (EOS) of nuclear matter
(see e.g. \cite{Speicher:1992,*Schmid:1995a} for early approaches and
\cite{Schunck:2019,*Meng:2016,*Lalazissis:2004} for recent reviews).
A suitable starting candidate seems to be the relativistic mean--field (RMF)
theory \cite{Miller:1972,Walecka:1974}. The RMF model has greatly
contributed to our understanding of nuclear structure, and it was
intensively used over years for a description of a wide class of
nuclear phenomena (see e.g. \cite{Bender:2003} and references therein
for reviews). Two mutually independent versions of the RMF
approach have been developed, either one with nonlinear
self--interactions of meson fields
\cite{Boguta:1977,Bodmer:1991,Gmuca:1992,Mueller:1996},
or one with density--dependent meson--nucleon couplings
\cite{Fuchs:1995,Typel:1999,Shen:1997,Hofmann:2001}. Both approaches cast
the additional density dependence into the effective interaction
that is inevitable for a proper description of nuclear systems.

The density--dependent couplings in the relativistic mean--field
approaches are introduced phenomenologically. They are often
determined by comparison with Dirac--Brueckner--Hartree--Fock (DBHF)
calculations for nuclear matter \cite{Brockmann:1990,Huber:1995,Sammarruca:2003,
vanDalen:2004,Katayama:2013} in limited range of densities.
Their functional forms are usually chosen {\it ad--hoc}; the choice being
dictated mainly by the requirement of simplicity.
Most authors employ the rational form.
Various functions of density dependences used are discussed
in Ref. \cite{Petrik:2012} together with their extrapolation
properties out of fitted interval of densities. Especially, the couplings
for low--density region (relevant for a nuclear periphery of exotic
nuclei) and high--density one (typical for neutron star calculations)
are not reliably determined by extrapolations.

Also, there exist approaches incorporating the exchange
(Fock) terms into the relativistic description of nuclear matter and
finite nuclei. The relativistic Dirac--Hartree--Fock (DHF) approach
\cite{Horowitz:1983,Bouyssy:1987} has been developed together with
its own density--dependent versions \cite{Fritz:1993,Shi:1995}.
The exchange parts give rise
to the state--dependent potentials due to nonlocal character of the
DHF approach and the solutions of Dirac equations for finite nuclei
thus require much more numerically intensive efforts than in the RMF
model.

To simplify the problem, the concept of equivalent local potentials
to the nonlocal exchange ones was intensively developed from early applications
of the DHF mean--field approach \cite{Miller:1972,Jaminon:1981}.
Next step was done in Refs.~\cite{Schmid:1995a,Schmid:1995b} where the exchange
energy density of the $\sigma$--$\omega$ model was evaluated analytically
in nuclear matter using the dilogarithm integrals, and its density dependence was
used to construct a local effective exchange potential.
Greco {\it et al.} \cite{Greco:2001} treated the Fock exchange term
in a ``kinetic approach''
employing the Wigner transform formalism. The approach was restricted to the
$\sigma$--$\omega$ model with the scalar self--interaction terms.

It is, therefore, of interest to develop an approach
as simple as the RMF model that will account for exchange
correlations, at least for nuclear matter \cite{Giai:2010}.

The present paper is aimed at mapping the Fock exchange terms of the
DHF energy density for nuclear matter onto its Hartree parts. An
explicit evaluation of the exchange integrals leads to the RMF
theory with density--dependent couplings. The analytical functional
density dependence of couplings obtained thus accounts for exchange
correlations over a full range of densities. This may significantly
reduce the arbitrariness of couplings at low and high densities.

For simplicity, we will treat the symmetric nuclear matter case and
consider the exchange of isoscalar mesons only, together with
$\pi$--meson. While the former gives the dominant contributions in
both, the RMF and DHF approaches, the pionic field contributes only
in an exchange part of the DHF model. This illustrates the essential
points of the approach. The generalization for asymmetric  nuclear
matter and the full set of exchanged mesons can be made
straightforwardly.

%
\section{DHF energy density for symmetric nuclear matter}
\label{sec:dhf2edf}

We start with the effective Lagrangian density for the interacting nucleons
($\psi$) and the scalar $\sigma$, vector $\omega$, and pion $\pi$ fields,
\begin{equation}
  {\mathcal L} = {\mathcal L}_0 + {\mathcal L}_I    \;,
\end{equation}
consisting of the free part
\begin{eqnarray}
  {\mathcal L}_0 & = & \overline{\psi}( i\gamma_\mu\partial^\mu - M)\psi
  + \frac{1}{2}(\partial_\mu\sigma\partial^\mu\sigma -m_{\sigma}^2\sigma^2)   \\
  & & -\frac{1}{4} \omega_{\mu\nu}\omega^{\mu\nu}
  + \frac{1}{2} m_{\omega}^2 \omega_\mu \omega^\mu
  + \frac{1}{2} (\partial_{\mu} \pi \partial^{\mu}\pi - m_{\pi}^2 \pi^2)
      \;,  \nonumber
\label{eq:Lag}
\end{eqnarray}
where $\omega_{\mu\nu} = \partial_{\mu}\omega_{\nu}-\partial_{\nu}\omega_{\mu}$,
and the interaction term
\begin{equation}
  {\mathcal L}_I = -g_{\sigma} \overline{\psi}\sigma\psi
  - g_{\omega} \overline{\psi}\gamma_{\mu} \omega^{\mu}\psi
  - \frac{f_{\pi}}{m_{\pi}} \overline{\psi} \gamma_5
 \gamma_{\mu} (\partial^{\mu} \pi)  \psi     \;.
\end{equation}
Here, the symbol $M$ denotes the nucleon rest mass, whereas
$g_i$ ($f_i$), $m_i$, $i=\{\sigma,\omega,\pi\}$
mean the coupling constants and rest masses for the scalar ($\sigma$),
vector ($\omega$) and pseudoscalar ($\pi$) mesons, respectively.
The direct Yukawa couplings are used for
the $\sigma$ and $\omega$ fields,
while the pseudovector coupling is applied for the pion.

To keep the paper self--explanatory we repeat here some basic facts
of the Dirac--Hartree--Fock approach for nuclear matter.
They are based mostly on Refs. \cite{Horowitz:1982} and
\cite{Bouyssy:1985}.

Due to parity and time--reversal symmetry, the nuclear matter
self--energy $\Sigma$ takes a form
\begin{equation}
 \Sigma(k) = \Sigma^s(k) + \gamma^0\Sigma^0(k)
 + \bm{\gamma} \cdot \bm{k} \Sigma^v(k) \; ,
\end{equation}
where superscripts $\{s,0,v\}$ denote the scalar, time--like and space--like
components of the self--energy, consecutively.

Dirac equation for nuclear matter is then written as
\begin{equation}
(\bm{\alpha} \cdot \bm{k^\star} + \beta M^\star) u(\bm{k},s) =
   E^\star u(\bm{k},s),
\end{equation}
with the positive energy solution being
\begin{equation}
u(\bm{k},s) = \left( \frac{E^\star + M^\star}{2 E^\star} \right)^{1/2}
              \left(
                 \begin{array}{c}
                     1 \\
                     \dfrac{\bm{\sigma} \cdot \bm{k^\star}}
                  {E^\star + M^\star}
                 \end{array}
              \right) \chi_s  \;.
\end{equation}
Here $\chi_s$ is a two--component Pauli spinor, and the starred
quantities are defined via the self--energy components as
\begin{subequations}
\begin{eqnarray}
  \bm{k^\star} = \bm{k}+\bm{k} \Sigma^v(k)  , \\
  M^\star = M + \Sigma^s(k)  ,                \\
  E^\star = E-\Sigma^0(k)  ,
\end{eqnarray}
\end{subequations}
and the on--shell condition is
\begin{equation}
 {E^\star}^2 = ({M^\star}^2 + {\bm{k^\star}}^2 )   \;.
\end{equation}
Additionally, the auxiliary functions $\hat{P}$ and $\hat{M}$
are introduced by relations
\begin{subequations}
\begin{eqnarray}
 \hat{P}(k) = k^\star / E^\star  , \\
  \hat{M}(k) = M^\star / E^\star  .
\end{eqnarray}
\end{subequations}

The energy functional, i.e., the nuclear matter energy density,
is given as (00)--component of the momentum--energy tensor.
Alternatively, it is obtained by taking the expectation value
of the Hamiltonian $\mathcal{H}$ with
respect to the ground state $\left| \Psi_0 \right>$ in a given volume
$\Omega$,
\begin{equation}
\label{eq:edf}
\mathcal{E} = \frac{1}{\Omega}
\left< \Psi_0 \right| \mathcal{H} \left| \Psi_0 \right>
\equiv
\left< \mathcal{T} \right> + \left< \mathcal{V} \right> ,
\end{equation}
and expressed as the sum of kinetic $\mathcal T$ and potential energy
$\mathcal V$ parts.

The potential energy density $\left< \mathcal{V} \right>$ of the DHF approach
can be decomposed into the direct (Hartree) term, $\varepsilon^D$,
and the exchange (Fock) contribution, $\varepsilon^X$,
\begin{equation}
\left< \mathcal{V} \right> = \varepsilon^D + \varepsilon^X   \;.
\end{equation}

%

The direct energy density reads
\begin{equation}
\varepsilon^D = -\frac{1}{2}\frac{g_{\sigma}^2}{m_{\sigma}^2}\rho _S^2
                +\frac{1}{2}\frac{g_{\omega}^2}{m_{\omega}^2}\rho _B^2   \;,
\end{equation}
where the scalar density $\rho_S$ and the vector (baryon) one $\rho_B$
were used. They are the sources of the corresponding meson--fields at
the Hartree (mean--field) level and are simply expressed via the nucleon
Fermi momentum $k_F$ in the nuclear matter as,
\begin{eqnarray}
\rho_B(k_F) &=& \frac{4}{(2 \pi)^3} \int_0^{k_F} d^3q =
                          \frac{4}{6 \pi^2} k_F^3 \;,\\
\rho_S(k_F,M^{\star}) &=& \frac{4}{(2 \pi)^3} \int_0^{k_F} d^3q
                          \frac{M^{\star}(q)}{E^{\star}(q)}  \;.
\end{eqnarray}
The pre--integral numerical factors of 4 reflect the spin/isospin degeneracy
of the nucleons in symmetric nuclear matter.
Only the $\sigma$ and $\omega$  mesons with direct
couplings contribute to $\varepsilon^D$. As it is already expressed
in terms of densities, it has the proper form for use in the DFT approach.

Variation of $\varepsilon^D$ with respect to the Dirac spinor
gives the direct contributions to the scalar, $\Sigma^s$, and time--like vector,
$\Sigma^0$, components of the self--energy.
One obtains,
\begin{equation}
  \Sigma^s = -\dfrac{g_\sigma^2}{m_\sigma^2} \rho_S ,   \qquad
 \Sigma^0 =  \dfrac{g_\omega^2}{m_\omega^2} \rho_B   \;.
\end{equation}
There is no contribution to space--like part of the vector component
of the self--energy.

%

In contrast to the direct (Hartree) energy density the exchange (Fock)
energy density, $\varepsilon^X$, has a more complicated structure.
All mesons contribute to the exchange energy density and it
does not exhibit the explicit dependence on nuclear densities:
\begin{equation}
    \varepsilon^X = \sum\limits_{\kappa} \varepsilon_{\kappa}^X ,
    \hspace{3em}  \kappa  =\{\sigma, \omega, \pi \} \nonumber
\end{equation}
\begin{eqnarray}
\label{eq:Xdensity}
    \varepsilon_{\kappa}^X & = & \frac{1}{(2\pi)^4}
    \int \limits_0^{k_F} k\, dk \int \limits_0^{k_F} q\, dq
    \left[  A_{\kappa} (k,q)   \right.         \\
    && \left. + \hat M(k) \hat M(q) B_{\kappa}(k,q)
    + \hat P(k) \hat P(q)  C_{\kappa}(k,q) \right]  \;, \nonumber
\end{eqnarray}
where $A_{\kappa}$, $B_{\kappa}$, $C_{\kappa}$ are known functions
for each meson exchanged and interaction type considered
(e.g. Ref.~\cite{Bouyssy:1987}); they are listed in Table~\ref{tab:DHFfactors}
for mesons involved in the present paper. The terms contain the angular exchange
integrals $\Theta_\kappa$ and $\Phi_\kappa$ for
$\kappa=\{\sigma, \omega, \pi\}$ given as
\begin{equation}
 \Theta_\kappa (k,q) = \ln \frac{(k+q)^2 + m_\kappa^2}{(k-q)^2 + m_\kappa^2} \;,
 \label{eq:Theta}
\end{equation}
and
\begin{equation}
 \Phi_\kappa (k,q) = \frac{1}{4 k q}(k^2+q^2+m_\kappa^2)\Theta_\kappa(k,q) - 1
      \;,
 \label{eq:Phi}
\end{equation}
where we already omitted the retardation effects in the meson propagators.

%
\begin{table}[ht]
\caption{The terms $A_{\kappa}$, $B_{\kappa}$, and $C_{\kappa}$ of
              Eq.~(\ref{eq:Xdensity})}
\label{tab:DHFfactors}
\begin{ruledtabular}
\begin{tabular}{cccc}
 $\kappa$ & $A_\kappa(k,q)$ & $B_\kappa(k,q)$ & $C_\kappa(k,q)$        \\
\hline  \noalign{\vskip 3pt}
$\sigma$   & $g_\sigma^2\Theta_\sigma$  & $g_\sigma^2\Theta_\sigma$  &
$-2g_\sigma^2\Phi_\sigma$  \\
$\omega$   & $2g_\omega^2\Theta_\omega$ & $-4g_\omega^2\Theta_\omega$&
$-4g_\omega^2\Phi_\omega$  \\
$\pi$      & $-3\left(\dfrac{f_\pi}{m_\pi}\right)^2m_\pi^2\Theta_\pi$
           & $-3\left(\dfrac{f_\pi}{m_\pi}\right)^2m_\pi^2\Theta_\pi$
           & $C_\pi$      \\
           & \multicolumn{3}{c}
 {$C_\pi= 6\left(\dfrac{f_\pi}{m_\pi}\right)^2\left[\left(k^2+q^2\right)\Phi_\pi
             - kq\Theta_\pi\right]$}  \\
\end{tabular}
\end{ruledtabular}
\end{table}
Nevertheless, using the approach developed in Appendix~\ref{sec:AppendixA},
we were able to evaluate all the exchange integrals in Eq.~(\ref{eq:Xdensity}).
Then with the results reached one may write down all contributions
to the exchange energy density in the closed--form.

Namely, for the $\sigma$--meson exchange we obtain
\begin{eqnarray}
\varepsilon_\sigma^X &=& \frac{1}{16} \frac{g^2_\sigma}{m^2_\sigma}
                X_{\theta}(k_F/m_\sigma) \rho_B^2 (k_F)        \nonumber  \\
                                &+& \frac{1}{16} \frac{g^2_\sigma}{m^2_\sigma}
                X_{\theta}(k_F/m_\sigma) \rho_S^2 (k_F,M^\star)            \\
                                &-& \frac{2}{16} \frac{g^2_\sigma}{m^2_\sigma}
                X_{\phi}(k_F/m_\sigma) \left(\rho_B^2 (k_F)
                                -\rho_S^2 (k_F,M^\star)\right)   \;.   \nonumber
\end{eqnarray}
The first line on the right-hand-side (rhs) of the last equation
comes from the evaluation of A--function term of Eq.~(\ref{eq:Xdensity}).
It depends on the square of vector (baryon) density $\rho_B$.
Additional density dependence 
\footnote{In fact, Fermi momentum dependence. However, as of
a definite relation between $k_F$ and $\rho_B$, we will often use
the terms \textquotedblleft Fermi momentum--dependent\textquotedblright 
and \textquotedblleft density--dependent\textquotedblright  interchangeably.}
is due to the exchange shape function $X_\theta$
via the dimensionless Fermi momentum variable $x=k_F/m_\sigma$.
After variation it contributes to the time--like vector component of
the self--energy.

The term on the second line of the rhs is from the evaluation of B--function
integral. It depends on the square of the scalar density. Similarly,
as in the previous case, the additional density dependence is determined
by the function $X_\theta (k_F/m_\sigma)$.
It gives contribution to the scalar component of the self--energy.

\begin{figure}[ht]
\includegraphics[width=75mm]{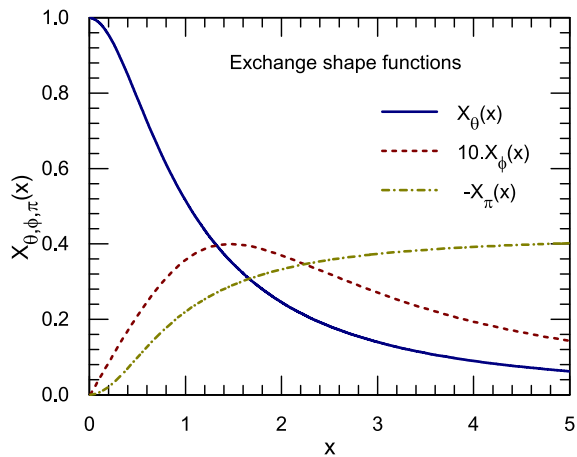}
\caption{\label{fig:X_functions}(Color online.)
The graph of the exchange shape
functions $X_\theta(x)$, $X_\phi(x)$, and $X_\pi(x)$.
The $X_\phi$ function is scaled by a factor of 10 for clarity,
and the sign of the function $X_\pi$ was changed to the opposite
one to keep the figure compact.}
\end{figure}

The last term on the third line has a different structure. It comes
from the integration of the C--function term. Firstly, the additional
density dependence is due to the function $X_\phi$. It behaves differently
in comparison with the function $X_\theta$. It vanishes at zero density and
grows very slowly with increasing Fermi momentum. The comparison of both,
the $X_\theta$ and $X_\phi$ exchange shape functions, is shown
in Fig.~\ref{fig:X_functions}. (Note that $X_\phi$ is scaled by a factor of 10
for better clarity.)
Secondly, the term depends on the difference of squares of the baryon and scalar
densities, respectively. Thus the term that in the DHF approach contributes
to the space--like vector component of the self--energy is split up to the scalar
and time--like vector components of the self--energy in the current procedure.
The DHF space--like component is thus being effectively eliminated.

%
%

%
For the $\omega$--meson exchange one writes
\begin{eqnarray}
\varepsilon_\omega^X &=& \frac{2}{16} \frac{g^2_\omega}{m^2_\omega}
             X_{\theta}(k_F/m_\omega) \rho_B^2 (k_F) \nonumber  \\
                                &-& \frac{4}{16} \frac{g^2_\omega}{m^2_\omega}
             X_{\theta}(k_F/m_\omega) \rho_S^2 (k_F,M^\star)  \\
                                &-& \frac{4}{16} \frac{g^2_\omega}{m^2_\omega}
             X_{\phi}(k_F/m_\omega) \left(\rho_B^2 (k_F)
                               -\rho_S^2 (k_F,M^\star) \right)   \;,  \nonumber
\end{eqnarray}
and the $\pi$--meson exchange contribution is
\begin{eqnarray}
\varepsilon_\pi^X &=& -\frac{3}{16} \left(\dfrac{f_\pi}{m_\pi}\right)^2
          X_{\theta}(k_F/m_\pi) \rho_B^2 (k_F)   \nonumber  \\
                          &-& \frac{3}{16}\left(\dfrac{f_\pi}{m_\pi}\right)^2
          X_{\theta}(k_F/m_\pi) \rho_S^2 (k_F,M^\star)  \\
                         &+& \frac{6}{16} \left(\dfrac{f_\pi}{m_\pi}\right)^2
          X_{\pi}(k_F/m_\pi) \left(\rho_B^2 (k_F)
                         -\rho_S^2 (k_F,M^\star) \right)  \;. \nonumber
\end{eqnarray}

Similar comments as were made to the $\varepsilon_\sigma^X$ part,
may be applied to the $\varepsilon_\omega^X $ and $\varepsilon_\pi^X$
contributions, except the $X_\pi$ shape function. It comes from
the $C_\pi$--term integration (see Appendix~\ref{sec:AppendixA})
and its graph is plotted in Fig.~\ref{fig:X_functions}. Due to smaller
rest mass of the pion, the dimensionless variable $x=k_F/m_\pi$ covers
a wider range of values than for other mesons, and thus produces a more
pronounced density dependence.

%

Now, combining the direct and exchange contributions and regrouping
them according to the scalar and vector densities, one may write the potential
energy density as
\begin{equation}
\left< \mathcal{V} \right> =
     -\frac{1}{2}\frac{\Gamma_{\sigma}^2(k_F)}{m_{\sigma}^2}\rho _S^2
    +\frac{1}{2}\frac{\Gamma_{\omega}^2(k_F)}{m_{\omega}^2}\rho _B^2 \;,
\end{equation}
which resembles the form of the mean--field (Hartree) contribution now, however,
with the effective density--dependent couplings $\Gamma_{\sigma}$ and
$\Gamma_{\omega}$. They are given by relations
\begin{widetext}
\begin{eqnarray}
\label{G2_sigma}
\frac{\Gamma_{\sigma}^2(k_F)}{m_{\sigma}^2} = \frac{g_{\sigma}^2}{m_{\sigma}^2}
&&-\frac{1}{8}\frac{g_{\sigma}^2}{m_{\sigma}^2}X_{\theta}(k_F/m_{\sigma})
+\frac{4}{8}\frac{g_{\omega}^2}{m_{\omega}^2}X_{\theta}(k_F/m_{\omega})
+\frac{3}{8}\left(\frac{f_{\pi}}{m_{\pi}}\right)^2 X_{\theta}(k_F/m_{\pi})  \\
&&
-\frac{2}{8}\frac{g_{\sigma}^2}{m_{\sigma}^2}X_{\phi}(k_F/m_{\sigma})
-\frac{4}{8}\frac{g_{\omega}^2}{m_{\omega}^2}X_{\phi}(k_F/m_{\omega})
+\frac{6}{8}\left(\frac{f_{\pi}}{m_{\pi}}\right)^2 X_{\pi}(k_F/m_{\pi})
\;,  \nonumber
\end{eqnarray}
and
\begin{eqnarray}
\label{G2_omega}
\frac{\Gamma_{\omega}^2(k_F)}{m_{\omega}^2} = \frac{g_{\omega}^2}{m_{\omega}^2}
&&+\frac{1}{8}\frac{g_{\sigma}^2}{m_{\sigma}^2}X_{\theta}(k_F/m_{\sigma})
+\frac{2}{8}\frac{g_{\omega}^2}{m_{\omega}^2}X_{\theta}(k_F/m_{\omega})
-\frac{3}{8}\left(\frac{f_{\pi}}{m_{\pi}}\right)^2 X_{\theta}(k_F/m_{\pi})   \\
&&
-\frac{2}{8}\frac{g_{\sigma}^2}{m_{\sigma}^2}X_{\phi}(k_F/m_{\sigma})
-\frac{4}{8}\frac{g_{\omega}^2}{m_{\omega}^2}X_{\phi}(k_F/m_{\omega})
+\frac{6}{8}\left(\frac{f_{\pi}}{m_{\pi}}\right)^2 X_{\pi}(k_F/m_{\pi})
\;,  \nonumber
\end{eqnarray}
\end{widetext}
%

%

These relations are the main results of the section. They represent the mean
field model with the density--dependent couplings that accounts for exchange
correlations from the DHF model in nuclear matter.
The approach is parameter--free in the sense that no new parameter was
introduced except the DHF ones already used. Actually, in this way we have
mapped the DHF model onto the RMF one.

\begin{figure}[ht]
\includegraphics[width=75mm]{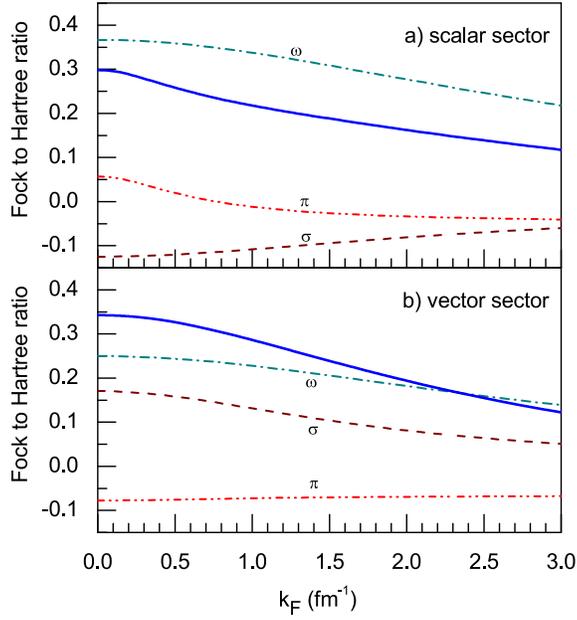}
\caption{\label{fig:F2H}(Color online.)
The ratios of Fock (exchange) to Hartree
(direct) parts of scalar (upper panel, Eq.(17))
and vector (lower panel, Eq.(15)) vertices, respectively.
The thick full line represents the result according to the Eq.(17),
resp. Eq.(15), while dashed, dot-dashed and dot-dot-dashed lines
are contributions from the exchange of particular mesons.
The thin solid curve is a result of a full numerical evaluation
of exchange integrals; it practically coincides with the analytical
result (thick solid line).}
\end{figure}

The approach was tested against the results of DHF calculations
\cite{Bouyssy:1987}.
Here, we have taken the DHF parameters of row (c) of Table~I there,
i.e. the parameterization of $\sigma$--$\omega$--$\pi$ DHF model.
The comparisons of the exchange contributions
to the effective  RMF couplings relative to the direct (Hartree) parts for both,
the scalar and vector channels, are shown in Fig.~\ref{fig:F2H}.

%

In both channels, the exchange correlations account for
$\sim$30\% of total coupling strength. This explains the
relations already observed in comparisons of the RMF and DHF
calculations; the results are similar provided the RMF couplings are
suitably renormalized.

The largest exchange contributions to both, the scalar and vector sectors,
come from the $\omega$--meson exchange. It contributes from $\sim$25\%
of the total coupling strength square for the vector sector up to more than
$\sim$35\% for the scalar channel at zero density and gradually decreases
with increasing Fermi momentum $k_F$.
The scalar meson exchange additions to the total coupling strengths
are significantly weaker than the $\omega$--meson ones for both, the scalar
and vector channels, respectively. It starts at $\sim-12$\% at zero density
for the scalar channel and slowly goes to zero for high--density matter.
At the vector sector, the scalar contribution starts at $\sim$18\% at zero
density and again slowly decreases with an increasing one.
The $\pi$--meson exchange contributions have a slightly different characters.
While the $\sigma$-- and $\omega$--exchanges are dominated by the A--term
and B--term integrals (due to the $X_\theta$ function behavior), leaving
the C--term less important (compare the $X_\phi$ function vs $X_\theta$ one),
the $\pi$ contribution depends on the combination of $X_\theta$ and
$X_\pi$ functions.
These contribute differently to the scalar and vector sectors, and due to small
$\pi$--meson mass, $m_\pi$, are of comparable strengths.
As a result, the $\pi$--meson addition to the effective scalar coupling square
starts at $\sim$6\% for zero density, decreases rapidly to $\sim$-5\% at
saturation and then remains roughly constant up to the dense matter region.

\begin{figure}[ht]
\includegraphics[width=75mm]{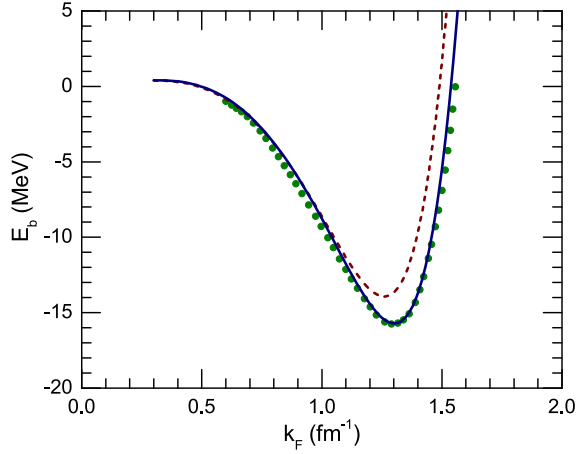}
\caption{\label{fig:Eb}(Color online.)
Binding energy per particle in symmetric nuclear matter.
The symbols correspond to the DHF results of Ref.~\cite{Bouyssy:1987},
and the solid line is the full RMF calculation with $k_F$--dependent
scalar and vector couplings. The dashed line represents the RMF result
with the space--like DHF contributions to the mapped RMF couplings omitted
(see text).}
\end{figure}

The final sensitive test of the couplings obtained  is made by comparison
of the original Bouyssy's results for the binding energy of nuclear matter
to our mean--field calculations with Fermi momentum dependent vertices
given by Eqs.~(\ref{G2_sigma},\ref{G2_omega}). This comparison is shown
in Fig.~\ref{fig:Eb}. The points are the results of the original
DHF calculations \cite{Bouyssy:1987}.
The solid line represents our mean--field calculations with Fermi momentum
dependent effective couplings including the direct and all exchange
contributions. It compares to  the original results fairly well.
The dashed line is the mean--field result similar to the solid
line, but omitting the contributions containing the $X_\phi$ and
$X_\pi$ exchange shape functions, i.e. omitting the contributions
from the space--like mapping on the mean--field approach.

The good agreement is further confirmed with a term--by--term comparison
of components of potential energies per nucleon at saturation of both,
the DHF calculation \cite{Bouyssy:1987} and our RMF model with the effective
density--dependent couplings. This is done in Table~\ref{tab:B_results}.

%
%

%
\begin{table}[ht]
\caption{\label{tab:B_results} Potential energies per particle at saturation}
\begin{ruledtabular}
\begin{tabular}{lrr}
             &  Ref.~\cite{Bouyssy:1987} &  this work             \\
\hline  \noalign{\vskip 3pt}
$\varepsilon^D_\sigma /\rho_B $  (MeV)   &  -171      & -176.5      \\
$\varepsilon^D_\omega /\rho_B $  (MeV)   &   145      &  145.1       \\
$\varepsilon^X_\sigma /\rho_B $  (MeV)   &    34      &    34.2         \\
$\varepsilon^X_\omega /\rho_B $  (MeV)   &    -24     &   -25.4       \\
$\varepsilon^X_\pi /\rho_B $  (MeV)      &    -6.7    &   -6.53        \\
\end{tabular}
\end{ruledtabular}
\end{table}

Thus we may conclude that the results obtained represent the mean--field model
with the density--dependent  (Fermi momentum--dependent) effective coupling
constants at the mean--field (Hartree) level that accounts for the exchange
(Fock) correlations.
This parameter--free approach may be easily cast on the full range of
exchanged mesons, as well as to consider a wider set of baryons. Also,
the asymmetric composition of nuclear matter may be taken into account.

%
\section{DHF with nonlinear self--couplings}
\label{sec:DHF_nl}

In the previous section, we have essentially reformulated
the linear Dirac--Hartree--Fock $\sigma$--$\omega$--$\pi$ model
for nuclear matter using the mean--field $\sigma$--$\omega$
approach with density--dependent effective couplings.
The density dependence (Fermi momentum dependence)
of the effective RMF couplings obtained due to contributions from
exchange of individual mesons is relatively weak. The
phenomenological density dependence of couplings in
relativistic mean--field calculations performed is usually stronger
\cite{Typel:2018}.
This indicates that other correlations, beyond the exchange (Fock) one,
are also playing important role in nuclear systems.
The field self--coupling or even cross--coupling terms are frequently employed
to account for the correlations mentioned \cite{Mueller:1996,Horowitz:2001}.

To take into account the effect of nonlinear self--interactions
we add the nonlinear terms for $\sigma$ and $\omega$ fields
into the Lagrangian (\ref{eq:Lag}), i.e.
\begin{equation}
U_\sigma = \frac{1}{3} b_2 (g_\sigma \sigma)^3
           + \frac{1}{4} b_3 (g_\sigma \sigma)^4      \;,
\end{equation}
and
\begin{equation}
U_\omega = \frac{1}{4} c_3 (g_\omega^2 \omega_\mu \omega^\mu)^2     \;,
\end{equation}
where the parameters $b_2$, $b_3$ represent the strength of cubic
and quartic self--couplings of the scalar field, while the parameter
$c_3$ is the quartic self--interaction strength of the vector field,
respectively.
Such self--interactions were studied in detail in the RMF approaches
with a considerable success ( see e.g. \cite{Boguta:1977,Bodmer:1991,
Gmuca:1992,Mueller:1996}, just to mention a few).

The simplest way to treat the self--couplings in the DHF model
is to introduce the effective meson masses and linearize the DHF
equation of motions \cite{Bernardos:1993}.
The equations for the respective $\sigma$ and $\omega$ fields then read
\begin{eqnarray}
  \partial^\mu \partial_\mu \sigma + {\hat{m}^{\star 2}_{\sigma}}\sigma
        & = & g_\sigma \overline{\psi} \psi   \; ,      \\
  \partial^\mu \partial_\mu \omega_\nu +  {\hat{m}^{\star 2}_{\omega}}\omega_\nu
       & = & g_\omega \overline{\psi} \gamma_\nu \psi   \; ,
\end{eqnarray}
where $\hat{m^\star}$ is the effective meson mass operator.
Then by replacing the fields with their expectation values and restricting
ourselves to nuclear matter we may write the relations for condensed fields
in the form
\begin{eqnarray}
  g_\sigma \sigma & = & - \dfrac{g^2_\sigma}{{m^\star_\sigma}^2} \rho_S   \;, \\
  g_\omega \omega & = & - \dfrac{g^2_\omega}{{m^\star_\omega}^2} \rho_S    \;,
\end{eqnarray}
where the respective effective masses are
\begin{equation}
  {m^\star_\sigma}^2 = m_\sigma^2 + b_2 \left( g_\sigma \sigma \right)
                                    +  b_2 \left( g_\sigma \sigma \right)^2     \;,
\end{equation}
and
\begin{equation}
  {m^\star_\omega}^2 = m_\omega^2 + c_3 \left( g_\omega \omega \right)^2   \;.
\end{equation}

Replacing the bare meson masses with their effective values reverts
the nonlinear DHF approach into its linear form, and subsequently
one may use the results of the previous section.

We, however, need to know the density dependence of effective masses.
This problem was solved in Appendix~\ref{sec:AppendixB}. With the results
obtained by solving the field equations in nuclear matter, we may write
the relation
\begin{equation}
  \dfrac{g^2_\kappa}{{m_\kappa^\star}^2} = \dfrac{g^2_\kappa}{m_\kappa^2}
         G_i (\beta_\kappa \rho_j) \;, \hspace{2em} \kappa=\{\sigma,\omega\},
\end{equation}
where the self--coupling shape functions $G_i$ depend on the selfinteraction
type, $\beta_\kappa$ represents the self--coupling strength, and $\rho_j$
denotes the source density for the given meson field, $j$=$S$ for the scalar
density, and $j$=$B$ for the vector density.

We were able to obtain very compact expressions for selfinteraction
shape functions for both, the cubic ($i$=2) and quartic ($i$=3) selfinteractions.
We were, however, unable to find a sufficiently well--looking expression
for cubic and quartic selfinteractions combined.
Therefore, if needed, it is better to use a combination of shape functions
for each selfinteraction alone.

%
\begin{figure}[t]
\includegraphics[width=75mm]{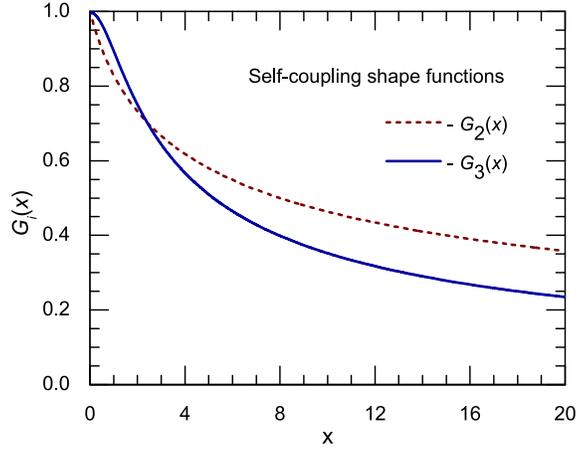}
\caption{\label{fig:G_functions}(Color online.)
The graph of the self--coupling shape
functions $G_i (x)$, $i$=2,3.}
\end{figure}
%

The shape functions obtained are
(see Appendix~\ref{sec:AppendixB})
\begin{equation}
 G_2 (x) = 2 \dfrac{\sqrt{1+x}-1}{x}
\end{equation}
for cubic self--coupling, and
\begin{equation}
 G_3(x) = \dfrac{3}{2} \dfrac{\sqrt[3]{\sqrt{1+x^2}+x}
            - \sqrt[3]{\sqrt{1+x^2}-x}}{x}
\end{equation}
for quartic selfinteractions. Both shape functions are normalized as
$G_i (0)=1$, for $i=\{2,3\}$, and the graphs of them are shown
in Fig.~\ref{fig:G_functions}.

%
\section{Applications and discussion}
\label{sec:Apps}

In preceding sections, we have mapped the DHF
energy density in nuclear matter onto RMF--like functional with
density--dependent effective couplings that are formally equivalent
to the DHF approach with nonlinear self--couplings.
The potential energy density of symmetric nuclear matter is finally
written as
\begin{equation}
  \left< \mathcal{V} \right> =
  - \frac{1}{2}\dfrac{\Gamma^2_\sigma}{m^2_\sigma} \rho^2_S
  + \frac{1}{2}\dfrac{\Gamma^2_\omega}{m^2_\omega} \rho^2_B  \;,
\end{equation}
where the density--dependent couplings now read
\begin{widetext}
\begin{eqnarray}
\frac{\Gamma_{\sigma}^2}{m_{\sigma}^2} =
\frac{g_{\sigma}^2}{m_{\sigma}^2} G_{3(2)}(\beta_\sigma \rho_S)
&&-\frac{1}{8}\frac{g_{\sigma}^2}{m_{\sigma}^2}G_{3(2)}(\beta_\sigma\rho_S)
       X_{\theta}(k_F/m_{\sigma})
+\frac{4}{8}\frac{g_{\omega}^2}{m_{\omega}^2}G_{3}(\beta_\omega \rho_B)
      X_{\theta}(k_F/m_{\omega})
+\frac{3}{8}\left(\frac{f_{\pi}}{m_{\pi}}\right)^2 X_{\theta}(k_F/m_{\pi})  \\
&&
-\frac{2}{8}\frac{g_{\sigma}^2}{m_{\sigma}^2}G_{3(2)}(\beta_\sigma \rho_S)
     X_{\phi}(k_F/m_{\sigma})
-\frac{4}{8}\frac{g_{\omega}^2}{m_{\omega}^2}G_{3}(\beta_\omega \rho_B)
    X_{\phi}(k_F/m_{\omega})
+\frac{6}{8}\left(\frac{f_{\pi}}{m_{\pi}}\right)^2 X_{\pi}(k_F/m_{\pi})
\;,  \nonumber
\end{eqnarray}
and
\begin{eqnarray}
\frac{\Gamma_{\omega}^2}{m_{\omega}^2} =
       \frac{g_{\omega}^2}{m_{\omega}^2}G_{3}(\beta_\omega \rho_B)
&&+\frac{1}{8}\frac{g_{\sigma}^2}{m_{\sigma}^2}G_{3(2)}(\beta_\sigma \rho_S)
          X_{\theta}(k_F/m_{\sigma})
+\frac{2}{8}\frac{g_{\omega}^2}{m_{\omega}^2}G_{3}(\beta_\omega \rho_B)
         X_{\theta}(k_F/m_{\omega})
-\frac{3}{8}\left(\frac{f_{\pi}}{m_{\pi}}\right)^2 X_{\theta}(k_F/m_{\pi})   \\
&&
-\frac{2}{8}\frac{g_{\sigma}^2}{m_{\sigma}^2}G_{3(2)}(\beta_\sigma \rho_S)
      X_{\phi}(k_F/m_{\sigma})
-\frac{4}{8}\frac{g_{\omega}^2}{m_{\omega}^2}G_{3}(\beta_\omega \rho_B)
      X_{\phi}(k_F/m_{\omega})
+\frac{6}{8}\left(\frac{f_{\pi}}{m_{\pi}}\right)^2 X_{\pi}(k_F/m_{\pi})
\;.  \nonumber
\end{eqnarray}
\end{widetext}

These expressions include both, the effect of Fock exchange via
the exchange shape functions $X_\theta$, $X_\phi$ and $X_\pi$,
and the medium modifications due to the nonlinear self--couplings
via the shape functions $G_i$.

A few simple cases should be mentioned:

\textbf{(a) Zero density constraints.}
The effective RMF couplings at zero density for both, the scalar and vector
channels are constrained by simple relations to the DHF bare couplings.
They are obtained by employing the values of shape functions at zero density.
One may write
\begin{equation}
 \left. \dfrac{\Gamma^2_\sigma}{m^2_\sigma} \right\rvert_{\rho_B=0}  =
  \frac{7}{8}\dfrac{g_{\sigma}^2}{m_{\sigma}^2}
+\frac{4}{8}\dfrac{g_{\omega}^2}{m_{\omega}^2}
+\frac{3}{8}\left(\dfrac{f_{\pi}}{m_{\pi}}\right)^2   \;,
\end{equation}
  and
\begin{equation}
  \left. \dfrac{\Gamma^2_\omega}{m^2_\omega} \right\rvert_{\rho_B=0} =
  \frac{10}{8}\dfrac{g_{\omega}^2}{m_{\omega}^2}
+\frac{1}{8}\dfrac{g_{\sigma}^2}{m_{\sigma}^2}
-\frac{3}{8}\left(\dfrac{f_{\pi}}{m_{\pi}}\right)^2   \;.
\end{equation}
These relations demonstrate an important contribution of the $\omega$--meson
exchange into both RMF effective couplings.

\textbf{(b) No density dependence of exchange.}
The density dependence due to exchange correlations is relatively weak.
Simple expressions for the effective couplings are obtained
when it is already omitted, i.e. the exchange shape functions are replaced
by their values at zero density, but keeping the self--interaction shape
functions fully density--dependent.
Then, the effective couplings read
\begin{equation}
  \dfrac{\Gamma^2_\sigma}{m^2_\sigma} =
  \frac{7}{8}\dfrac{g_{\sigma}^2}{m_{\sigma}^2} G_{3(2)}(\beta_\sigma \rho_S)
+\frac{4}{8}\dfrac{g_{\omega}^2}{m_{\omega}^2} G_3(\beta_\omega \rho_B)
+\frac{3}{8}\left(\dfrac{f_{\pi}}{m_{\pi}}\right)^2   ,
\end{equation}
and
\begin{equation}
  \dfrac{\Gamma^2_\omega}{m^2_\omega} =
   \frac{10}{8}\dfrac{g_{\omega}^2}{m_{\omega}^2} G_3(\beta_\omega \rho_B)
  +\frac{1}{8}\dfrac{g_{\sigma}^2}{m_{\sigma}^2} G_{3(2)}(\beta_\sigma \rho_S)
  -\frac{3}{8}\left(\dfrac{f_{\pi}}{m_{\pi}}\right)^2   .
\end{equation}
It is expected that the residual exchange density dependence is absorbed
into the self--interaction density dependence parameters.

This model is close to the earlier developed density--dependent approaches
(see e.g. \cite{Fuchs:1995,Typel:1999,Shen:1997,Hofmann:2001,Typel:2018}
and references therein). Now, however, the compositions of effective
mean--field couplings are dictated by the structure of exchange contributions,
and the density dependence is induced by the nonlinear self--interactions.

\textbf{(c) Low density limit.}
Previous results may be simplified further when the nuclear density
of the system is relatively low. In such a case, the scalar density may be
approximated by a scaled baryon density, i.e. $\rho_S \sim \rho_B$.
Then the scalar density dependencies are replaced by the vector densities
and the effective couplings take particularly simple forms,
\begin{equation}
  \dfrac{\Gamma^2_\sigma}{m^2_\sigma} =
  \frac{7}{8}\dfrac{g_{\sigma}^2}{m_{\sigma}^2}
      G_{3(2)}(\beta^\prime_\sigma \rho_B)
 +\frac{4}{8}\dfrac{g_{\omega}^2}{m_{\omega}^2}
      G_3(\beta_\omega \rho_B)
 +\frac{3}{8}\left(\dfrac{f_{\pi}}{m_{\pi}}\right)^2   ,
\end{equation}
and
\begin{equation}
  \dfrac{\Gamma^2_\omega}{m^2_\omega} =
   \frac{10}{8}\dfrac{g_{\omega}^2}{m_{\omega}^2}
       G_3(\beta_\omega \rho_B)
   +\frac{1}{8}\dfrac{g_{\sigma}^2}{m_{\sigma}^2}
       G_{3(2)}(\beta^\prime_\sigma \rho_B)
   -\frac{3}{8}\left(\dfrac{f_{\pi}}{m_{\pi}}\right)^2   .
\end{equation}

\begin{figure}[ht]
\includegraphics[width=75mm]{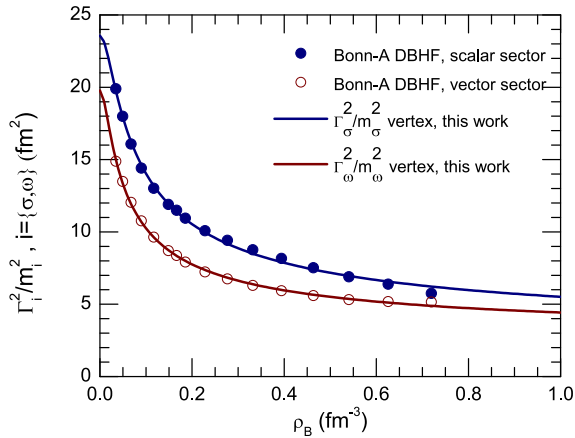}
\caption{\label{Fig:Effective_couplings}(Color online.)
Density--dependent effective couplings. The solid and opens symbols
were deduced from DBHF scalar and vector potentials, respectively,
of the Brockmann and Machleidt calculations \cite{Brockmann:1990}
for symmetric nuclear matter with Bonn A NN interaction.
The solid lines are the fits employing the density--dependent
self--coupling shape functions.}
\end{figure}

The model was used to fit the density--dependent vertices deduced from
the Dirac--Brueckner--Hartree--Fock (DBHF) calculations of nuclear matter
\cite{Brockmann:1990} with Bonn A interaction.
The DHF coupling--to--mass ratios, $g^2_\sigma/m^2_\sigma$ and
$g^2_\omega/m^2_\omega$, and the $\beta$'s were taken as free
parameters of the fits. The DBHF vertices and the effective RMF fitting curves
are compared in Fig.~\ref{Fig:Effective_couplings}.
The agreement is very good over the wide density region fitted despite
the fact that only baryon density dependence of both effective couplings
was used.

%
%

%
\section{Summary and conclusions}
\label{sec:Summary}

In the first part of the present work, we have mapped out the exchange Fock
contributions of the DHF approach for symmetric nuclear matter onto the
direct Hartree terms. This results in the effective relativistic mean--field
model with the density--dependent (Fermi momentum--dependent) couplings.
The density dependence of the coupling constants reflects the exchange
correlations.

The approach is based upon the evaluation of exchange integrals of the DHF
energy density components. All of them are expressed in the closed form
via relevant densities and introduce no new parameters. Thus, it presents
a step towards the formulation of the nonlocal Dirac--Hartree--Fock model
as the local density functional theory. The exchange shape functions
were derived that govern the $k_F$ dependence of Fock contributions.

The approach was tested against the DHF calculation of nuclear matter.
It was shown that the effective RMF model reproduces well the DHF results
for binding energy.

Alternatively, the approach may be developed by evaluating the exchange
integrals in the relations for DHF self--energy components \cite{Gmuca:2019}.
However, proceeding in this way we need to eliminate the momentum dependence
of the exchange terms that is an inherent feature of the relativistic
Hartree--Fock model.
That was done by averaging the terms over the Fermi sphere.
Contrary, no such approximations were introduced in the present work
starting from exchange energy densities.
Simply the $k$--dependence is internally integrated out.

It was observed that the density dependence of Fock exchange terms
is relatively weak. Therefore, the nonlinear self--coupling terms were added
to the DHF model.  This leads to an additional density dependence of
the effective RMF vertices that is decoupled from the Fock ones.

In the standard RMF model alone the scalar and vector self--interactions
contribute only to the corresponding scalar or vector channels, respectively.
The density dependence of the respective channel is thus fully determined
by the given self--interaction only, and governed by the phenomenological
ad--hoc parameterizations.

The model presented in this work exhibits a more complex density dependence.
The scalar and vector self--interactions contribute coherently to the density
dependence of both, the scalar and vector channels, together with
a small tuning due to the exchange density dependencies.

While this work is restricted to the exchange of $\sigma$--, $\omega$--,
and $\pi$--mesons  only, the approach itself can be straightforwardly
generalized for the full set of mesons,  and for asymmetric nuclear matter.
The contribution of this work is, we believe, a step in a direction to formulate
the effective nuclear density functional suitable for the description of a wide
class of nuclear phenomena over a broad range of densities. Such a development
is of utmost importance since the theory is needed able to predict nuclear
properties in regions of the nuclear chart not yet accessible to experiment.
It seems that the relativistic self--consistent mean--field model with
the density--dependent couplings, the structure and density dependence of which
come from the nonlinear DHF theory, is a good approach to reach this goal.

\begin{acknowledgments}
This work was supported in part by the VEGA Grant Agency under
project No. 2/0181/21 and by the Slovak Research and Development
Agency under contract No. APVV-15-0225,
and the JINR theme No. 02-1-1087-2009/2023.
\end{acknowledgments}


\appendix

\section{Evaluation of exchange integrals}
\label{sec:AppendixA}

By inspecting the DHF expression for the exchange energy density,
Eq.~(\ref{eq:Xdensity}),
one reveals that it depends crucially only on a few kinds of
exchange integrals without an apparent density dependence.
Usually, they were evaluated numerically.
%

In this section we will give a detailed procedure for evaluating
all types of exchange integrals of the DHF approach. The goal is
to express them explicitly in the form of density--dependent terms.

\subsection{The $A$--term integrals}

The basic exchange integral over $A$--function dependent terms reads
\begin{equation}
I^A = \frac{g^2}{(2\pi)^4}  \int_0^{k_F} k\, dk \int_0^{k_F}
          \Theta(k,q)\, q\, dq  \;,
\end{equation}
where $k_F$ is the nucleon Fermi momentum in nuclear matter, and the function
$\Theta(k,q)$ is written as (see Eq.~(\ref{eq:Theta}))
\begin{equation}
\Theta(k,q) = \ln \frac{(k + q )^2 + m^2}{(k - q )^2 + m^2} \;. \nonumber
\end{equation}
It depends on the mass of exchanged meson $m$ and
the square of meson coupling $g^2$.
Other numerical multiplicative factors were omitted
to keep the integral as simple as possible.

This integral may be evaluated in the closed form. After some algebra,
by employing the expression for a nucleon vector (baryon) density,
\begin{equation}
\rho_B (k_F) = \frac{2}{3 \pi^2} k_F^3 \;, \nonumber
\end{equation}
the integral may be finally written in a compact form as
\begin{equation}
I^A = \frac{1}{16} \frac{g^2}{m^2} X_{\theta}(k_F/m) \rho_B^2 (k_F) \; ,
\label{eq:A2}
\end{equation}
where the auxiliary exchange shape function $X_{\theta}(x)$ is introduced.
It reads
\begin{eqnarray}
X_{\theta}(x) &=& \frac{3}{32} \left[4x^2\left(6x^2-1-8x\arctan2x \right)
         \right.  \nonumber  \\
&& +\left.\left(12x^2+1\right)\ln(1+4x^2)\right]/x^6  \; .
\label{eq:A3}
\end{eqnarray}
Here the $X_{\theta}(0)=1$ normalization was chosen.
The graph of the function is plotted in Fig.~\ref{fig:X_functions}.

\subsection{The $B$--function integrals}

The $B$--function exchange integrals are similar as
the $A$--function ones, except they contains the $\hat{M}$ quantities.
Namely, the integral take the basic form
\begin{equation}
I^B = \frac{g^2}{(2\pi)^4} \int_0^{k_F} \hat{M}(k)k \,dk \int_0^{k_F} \hat{M}(q)
          \Theta(k,q) q \,dq \; ,
\label{eq:A4}
\end{equation}
where
\begin{equation}
\hat{M}(k) = \frac{M^{*}(k)}{E^{*}(k)}    \; ,       \nonumber
\end{equation}
and the quantities $M^{*}(k)$, $E^{*}(k)$ are the $k$--dependent
effective mass and Fermi energy, respectively.

To proceed further, we firstly rewrite the $A$--function integral
(A1) in an equivalent form as
\begin{equation}
I^A = \dfrac{g^2}{16} \int_0^{k_F} \dfrac{2}{\pi^2} k^2 dk \int_0^{k_F}
         \left\{\frac{\Theta(k,q)}{4 k q}\right\} \dfrac{2}{\pi^2} q^2 dq  \; .
\end{equation}
Then, by applying the mean value theorem, the integral may be rewritten as
\begin{equation}
I^A = \frac{g^2}{16} K \int_0^{k_F} \dfrac{2}{\pi^2}k^2 dk
              \int_0^{k_F}  \dfrac{2}{\pi^2}q^2 dq  \; ,
\end{equation}
where
\begin{equation}
  K = \left. \frac{\Theta(k,q)}{4 k q} \right|_{k=\overline{s},
        q=\overline{s}} \; ,
\end{equation}
is evaluated at some intermediate point $\overline{s}$ of the integrating
interval $\left( 0, k_F \right>$.
The remaining double integral results to the square of baryon density,
$\rho_B$. The actual value of the quantity $K$ is then obtained by comparison
with the relation (\ref{eq:A2}).

Then, by applying this procedure for an evaluation of the $B$--function
exchange integrals (\ref{eq:A4}), one finally obtains the resulting expression,
\begin{equation}
I^B = \frac{1}{16} \frac{g^2}{m^2} X_{\theta}(k_F/m) \rho_S^2 (k_F,M^{\star})
      \; ,
\label{eq:A8}
\end{equation}
where $\rho_S$ is the scalar density and $M^{\star}$ is the effective
nucleon mass.

\subsection{The $C$--function integrals}

The exchange integrals containing the $C$--term vertex functions are a bit
more complicated than in previous cases.
The form of integral depends upon the type of interaction.

For $\sigma$ and $\omega$ meson fields with direct (Yukawa) couplings
to the nucleon field the basic integral is
\begin{equation}
 I^C = \frac{g^2}{(2\pi)^4} \int_0^{k_F} \hat{P}(k)k \,dk
          \int_0^{k_F} \hat{P}(q)  \Phi(k,q) q \,dq \; ,
\label{eq:A9}
\end{equation}
where $\Phi(k,q)$ is an angular exchange integral (see Eq. (\ref{eq:Phi}),
\begin{equation}
 \Phi (k,q) = \frac{1}{4 k q}(k^2+q^2+m^2)\Theta (k,q) - 1 \;,  \nonumber
\end{equation}
and
\begin{equation}
  \hat{P}(k) = \frac{k^{\star}(k)}{E^{\star}(k)} \; .  \nonumber
\end{equation}

To evaluate the $I^C$ integral we will employ the techniques used for
the A-- and B--functions cases.
Firstly, we set $\hat{P}(k) \equiv 1$ in the integral (\ref{eq:A9}).
In this case the integral can be evaluated in the closed form,
\begin{equation}
  \left. I^C \right|_{\hat{P}\equiv 1}  = \frac{1}{16} \frac{g^2}{m^2}
         X_{\phi}(k_F/m) \rho_B^2 (k_F) \; ,
\end{equation}
where the exchange shape function $X_\phi$ is,
\begin{eqnarray}
&&X_{\phi}(x) = \frac{-3}{16} \left[2x^4+\left(x^2+1\right) \left(16x \arctan x
    - 8 \arctan 2x \right. \right.      \nonumber  \\
&& \left. \left.  +4(x^2-1) \ln (1+x^2)+(1-4x^2) \ln (1+4x^2)\right) \right]
      /x^6     \;.     \nonumber  \\
&&
\end{eqnarray}

Subsequently, using the mean value theorem (as in the case of B--term integrals)
and employing the on--shell condition, the C--term integral may be finally
expressed as,
\begin{equation}
  I^C = \frac{1}{16} \frac{g^2}{m^2} X_{\phi}(k_F/m)
            \left( \rho_B^2 (k_F)-\rho_S^2(k_F,M^\star)\right) \, .
\end{equation}
%

The C--type function for the pion with a pseudovector coupling is given by
a different expression (see Table~\ref{tab:DHFfactors}),
\begin{equation}
  C_\pi (k,q)= 6\left(\dfrac{f_\pi}{m_\pi}\right)^2\left[\left(k^2+q^2\right)
               \Phi_\pi - kq\Theta_\pi\right]   \; .   \nonumber
\end{equation}
The exchange integral $I^C_\pi$ to be evaluated is
\begin{equation}
  I^C_\pi = \dfrac{1}{(2\pi)^4} \int_0^{k_F} k\, dk \int_0^{k_F}
          C_\pi (k,q) q\, dq \; .
\end{equation}

Proceed in an analogical way as in the previous case we write
the $I^C_\pi$ integral in the closed form
\begin{equation}
  I^C_\pi = \frac{6}{16} \left(\dfrac{f_\pi}{m_\pi}\right)^2 X_{\pi}(k_F/m_\pi)
                 \left(\rho_B^2 (k_F) - \rho_S^2 (k_F,M^\star) \right)   \; ,
\end{equation}
where the exchange shape fubction $X_\pi$ is written as
\begin{eqnarray}
&&
X_{\pi}(x) = \frac{-1}{80} \left[ 56 x^6+18 x^4+\left( 40x^3-24x \right)
     \left( 2 \arctan x  \right. \right.      \nonumber  \\
&& \left. \left.
     - \arctan 2x \right)  - \left(16x^6 +60x^4 +5\right) \left( \ln(1+4x^2)
     \right. \right.  \nonumber   \\
&&  \left.  \left.
      -\ln(1+x^2) \right)
     +15\ln(1+x^2) \right] /x^6   \; .
\end{eqnarray}
The graphs of both shape functions, $X_\phi$ and $X_\pi$, are also shown in
Fig.~\ref{fig:X_functions}.

\section{Self--coupling shape functions}
\label{sec:AppendixB}

The meson field equations reduce in nuclear matter to simple algebraic equations,
\begin{equation}
  u = a (\rho - c u^n )  \;,   \quad     n=\{2,3\}  \; ,
\label{eq:B1}
\end{equation}
where $u$ is a meson potential, $\rho$ is the source density, and
the real constants $a$ and $c$ mean the coupling and self--coupling strengths.
The exponent $n$ takes the value 2 or 3, depending upon the cubic or quartic
self--coupling terms in the model Lagrangian density, respectively.
The solutions of (\ref{eq:B1}) are usually obtained in
a direct way or iteratively.

For the approach developed in this paper it is of advantage to express
the solutions in the form
\begin{equation}
  u = a\, G(\beta \rho)\, \rho   \; ,
\end{equation}
which resembles the solution of linear model.
We are interested in real solutions with the normalization $G(0)=1$.
The effect of self--couplings is hidden within the density--dependence of
the self--coupling shape function $G$, and the factor $\beta$ is the new
self--coupling strength. It can be expressed using the constants
$a$ and $c$ of (\ref{eq:B1}) (e.g. $\beta=4a^2c$ for the cubic self--interaction,
and $\beta=\sqrt{27 a^3 c}/2$ for quartic one, respectively).
We expect, however, that in most cases the values of $\beta$'s to be determined
in the model parameter optimization procedure.

The explicit form of the shape function is
\begin{equation}
  G_2(x) = 2 \dfrac{\sqrt{1+x} - 1}{x}  \; ,
\end{equation}
for the cubic self--interaction, and
\begin{equation}
 G_3(x) =\dfrac{3}{2} \dfrac{\sqrt[3]{\sqrt{1+x^2}+x} -
               \sqrt[3]{\sqrt{1+x^2}-x}}{x}  \;,
\end{equation}
for the quartic self--coupling.
The shape function graphs are shown in  Fig.~\ref{fig:G_functions}.


\providecommand{\noopsort}[1]{}\providecommand{\singleletter}[1]{#1}%
\end{document}